\documentclass[aps,prb,singlecolumn,groupedaddress]{revtex4-2}
\usepackage{bm}
\usepackage{graphicx}
\usepackage{color,soul}
\usepackage[T1]{fontenc}
\usepackage{ae,aecompl}
\usepackage{textgreek}
\usepackage{xcolor}
\usepackage{hyperref}

\begin{document}

\title[Comment on ``$\Phi$~memristor: Real memristor found'']{Comment on ``$\Phi$ memristor: Real memristor found'' by
F.~Z.~Wang, L.~Li, L.~Shi, H.~Wu, and L.~O.~Chua
[J. Appl. Phys. {\bf 125}, 054504 (2019)]}
\author{Y.~V.~Pershin}
\email{pershin@physics.sc.edu}
\address{Department of Physics and Astronomy, University of South Carolina, Columbia, South Carolina
29208, USA}

\author{M.~Di Ventra}
\email{diventra@physics.ucsd.edu}
\address{Department of Physics, University of California San Diego, La Jolla, California 92093, USA}

\begin{abstract}
Wang {\it et al.} claim [J. Appl. Phys. {\bf 125}, 054504 (2019)] that a current-carrying wire interacting with a magnetic core represents a memristor. Here, we demonstrate that this claim is false. We first show that such memristor ``discovery'' is based on incorrect physics, which does not even capture basic properties of magnetic core materials, such as their magnetic hysteresis. Moreover, the predictions of Wang {\it et al.}'s model contradict the experimental curves presented in their paper. Additionally, the theoretical pinched hysteresis loops presented by Wang {\it et al.} can not be reproduced if their model is used, and there are serious flaws in their ``negative memristor'' emulator design. Finally, a simple gedanken experiment shows that the proposed $\Phi$-memristor would fail the memristor test we recently suggested in J. Phys. D: Appl. Phys. {\bf 52}, 01LT01 (2019). The device ``discovered'' by Wang {\it et al.} is just an inductor
with memory.

\vspace{0.3cm}

\textcolor{blue}{Added on 01/15/2021. Following the present comment, the article [F.~Z.~Wang, L.~Li, L.~Shi, H.~Wu, and L.~O.~Chua,
J. Appl. Phys. {\bf 125}, 054504 (2019)] was retracted from JAP on technical grounds. The retraction notice~\cite{Retraction21a} can be found at \href{https://doi.org/10.1063/5.0040852}{https://doi.org/10.1063/5.0040852}.}
\end{abstract}

\maketitle

In a recent paper~\cite{Wang19a}, Wang {\it et al.} claimed the ``discovery'' of what they call the ``$\Phi$-memristor'' or ``real memristor'', in which a current-carrying wire
is strung through a magnetic core. As we show below, this claim is based on an {\it erroneous} physical model of magnetization dynamics that links
the normalized magnetization $m(t)$  to the net charge $q(t)$ that traverses the device according to
\begin{equation}
  m(t)=\textnormal{tanh}\left[ \frac{q(t)}{S_W} +C \right]  . \label{eq:1}
\end{equation}
where $S_W$ is a switching constant, $C=\textnormal{tanh}^{-1}m_0$ is a constant of integration, and $m_0$ is the initial magnetization. Based on Eq. (\ref{eq:1}), the authors of~\cite{Wang19a} derived the current-voltage relation of the ``real memristor'' of the form
\begin{equation}\label{eq:2}
  V(t)=\frac{\textnormal{d}\varphi}{\textnormal{d}t}=\frac{\mu_0SM_S}{S_W}\textnormal{sech}^2\left(\frac{q(t)}{S_W}+\textnormal{tanh}^{-1}m_0 \right)I(t).
\end{equation}
Here, $S$ is an area, and $M_S$ is the saturation magnetization. While, superficially, Eq.~(\ref{eq:2}) has the form of a memristor model, $V=R_M(q)I$~\cite{chua71a}, the fact that Eq.~(\ref{eq:2}) was derived based on incorrect physics refutes their claim.

Curiously, in Ref.~\cite{diventra09a} exactly the same system was given as an example of inductor with memory (meminductive system) by the authors of this Comment together with the last author of Ref.~\cite{Wang19a} (Chua). We continue to argue that it was the correct classification,
namely that a current-carrying wire strung through a magnetic core is {\it not} a memristor, but simply an inductor with memory.

\begin{figure*}[t]
 \centering (a) \includegraphics[width=45mm]{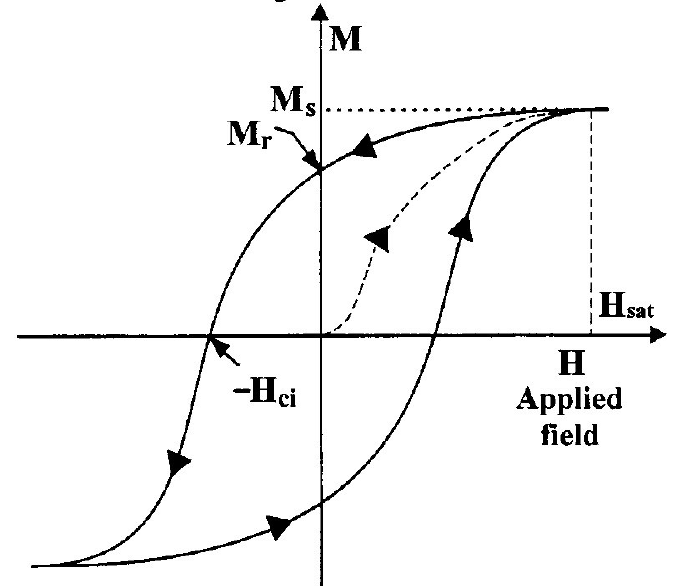}(b) \includegraphics[width=45mm]{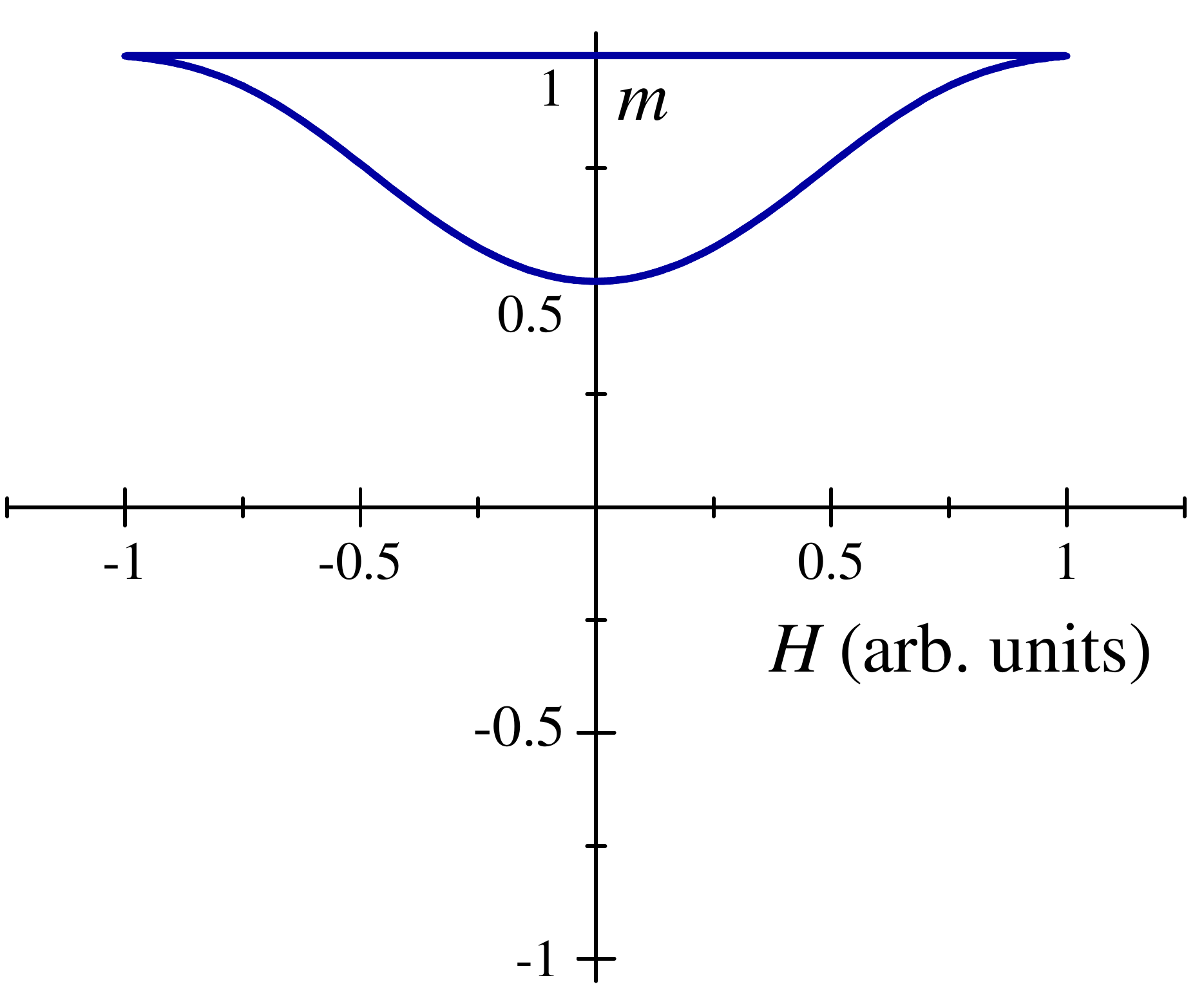}(c) \includegraphics[width=45mm]{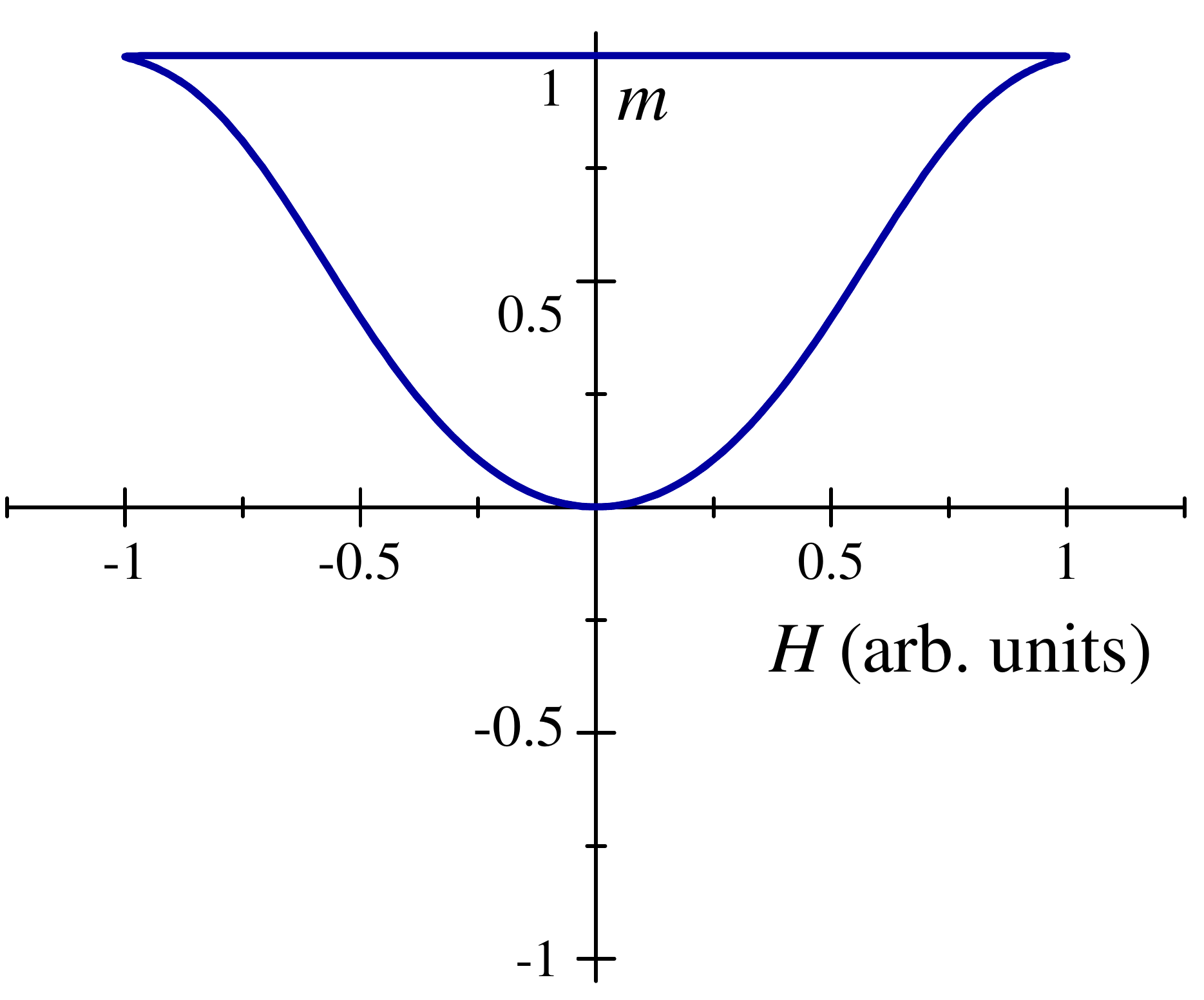}
  \caption{(a) Magnetization, $M$, vs magnetic field, $H$, curve for a ferromagnetic/ferrimagnetic material. Reprinted with permission from Ref.~\cite{Rudowicz03a}. (b) and (c): Normalized magnetization, $m$, versus $H$ plotted using Eq.~(\ref{eq:1}) for  $I(t)=I_0 \sin(\omega t)$ with parameter values $I_0/(\omega S_W)=10/3$, $m_0=0.5$ (b), and $m_0=0$ (c). \label{fig1}}
\end{figure*}

Typically, magnetic core memories utilize rings of semi-hard ferrite, which is a ferrimagnet. Ferrimagnets
are characterized by hysteretic magnetization loops, similar to the loops of ferromagnets (see, e.g., Refs.~\cite{Rudowicz03a} and \cite{Katz69a}). In Fig.~\ref{fig1}(a) we provide a typical magnetization curve
of ferromagnet/ferrimagnet materials showing that the magnetization switches between two limiting values ($\pm M_S$) in response to a time-dependent (e.g., sinusoidal) input. However, this property is missing in the Eq.~(\ref{eq:1}) model.
According to  Fig.~\ref{fig1}(b) and (c), the hysteretic loops obtained from Eq. (\ref{eq:1}) are not even close to the ones of real ferromagnets/ferrimagnets.
In fact, a general property of Eq.~(\ref{eq:1}) for any frequency of sinusoidal input is that the magnetization never decreases below its initial value.
Therefore, there is a clear {\it disagreement} between the response of actual physical ferromagnets/ferrimagnets and the dynamics described by Eq. (\ref{eq:1}).

\begin{figure*}[b]
  \centering (a)\includegraphics[width=60mm]{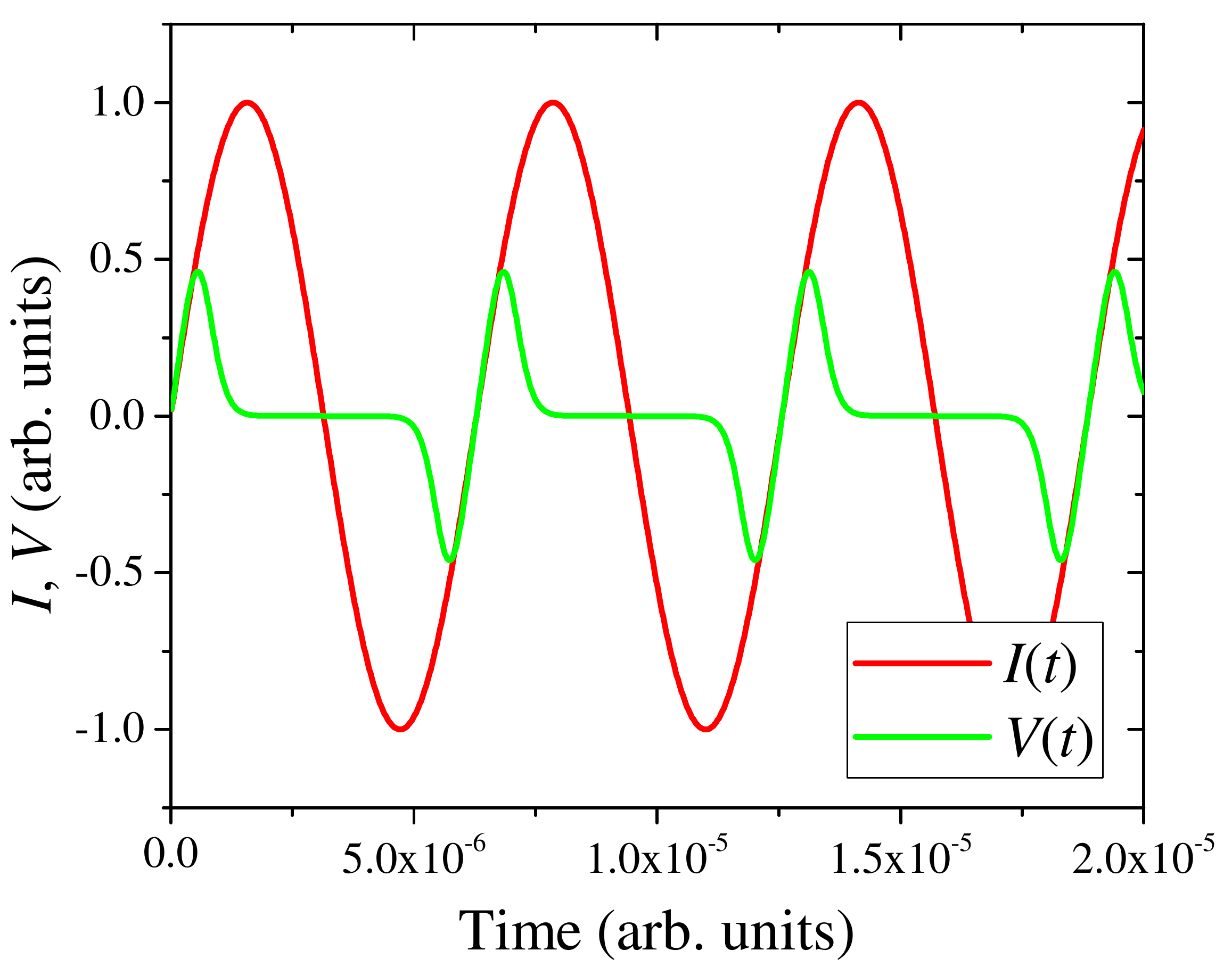} \hspace{0.5cm} (b)\includegraphics[width=60mm]{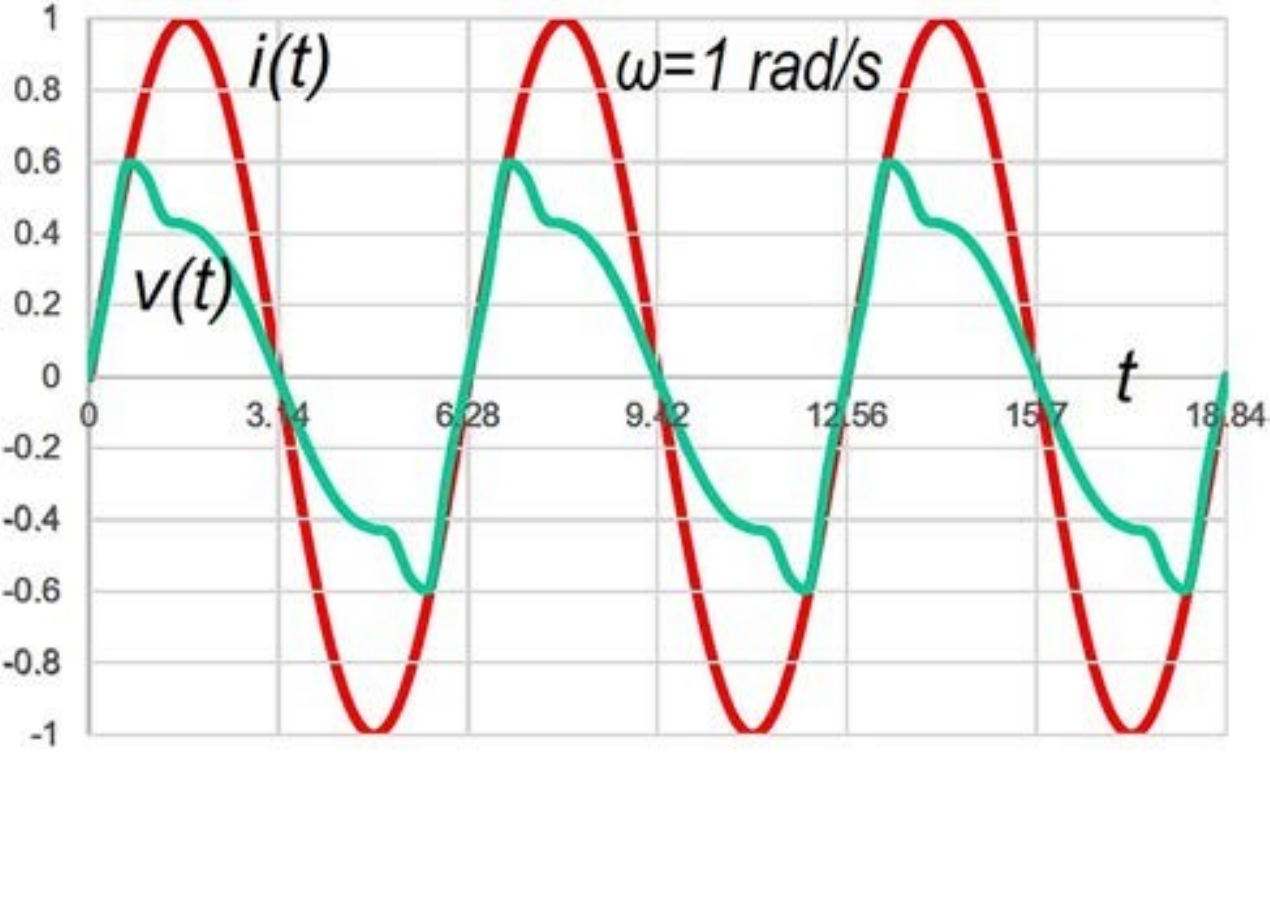}
  \caption{ (a) Time-dependencies of the current (red curve) and voltage (green curve). Here, $V(t)$ is  plotted using Eq.~(\ref{eq:2}) with a sinusoidal current input $I(t)=I_0 \sin(\omega t)$ for $I_0/(\omega S_W)=10/3$, $m_0=0$. (b) Corresponding plot reprinted with permission from Ref.~\cite{Wang19a}.}\label{fig2}
\end{figure*}

Next, in Fig.~\ref{fig2} we plot the voltage across the ``$\Phi$-memristor'' as a function of time. We emphasize that in this plot the voltage peaks representing the magnetization reversal have the {\it wrong timing}. While in real  materials one should expect 180$^{\circ}$ phase shift
between negative and positive peaks~\footnote{The experimentally-measured voltage response presented in Fig. 10 of Ref.~\cite{Wang19a} has the correct 180$^{\circ}$ phase shift, albeit for a different input waveform.}, the phase shift in Fig.~\ref{fig2}(a) is much longer.
Additionally, it should be mentioned that Fig.~5 of  Ref.~\cite{Wang19a} {\it can not be reproduced} using the analytical expression for the voltage across the memristor, Eq.~(\ref{eq:2}): compare Fig.~\ref{fig2}(a) and (b) that are supposed to represent the same responses. It seems that an undocumented sinusoidal
input was added to make $V(t)$ look closer to typical memristor curves found in the literature. There must also be a mistake with the frequency range used
by the authors of Ref.~\cite{Wang19a} since the magnetization-switching processes, which typically occur at the microsecond time scale, would be seen instantaneous on the scale of seconds used in Fig.~\ref{fig2}(b).

We further comment that the design of the ``negative memristor'' emulator (Fig. 19 of Ref.~\cite{Wang19a}), as it stands, is incomplete. It looks very similar to the digital memristor emulator suggested by us in Ref.~\cite{pershin10a}, but misses the connections between the measurement terminals and the switching resistors. Additionally, it is not obvious why the switching resistors are negative. Generally, negative resistances must be emulated by active components, such as digital-to-analog converters (DACs), see Ref.~\cite{Kolka12a}.

Finally, we provide yet another argument as to why the ``$\Phi$-memristor'' proposed by Wang {\it et al.} is not a memristor. Very recently, we have introduced a test that can unambiguously distinguish whether the tested device is indeed a memristor or something else~\cite{pershin18a}. The test is based on a duality property of the capacitor-memristor circuit, and can be summarized as follows: ``For any initial resistance states of the memristor and any form of
the applied voltage, the final state of an ideal memristor must be identical to its initial state,
if the capacitor charge finally returns to its initial value''~\cite{pershin18a}.

It is very easy to see that the ``$\Phi$-memristor'' would fail the above test by simply examining the following gedanken experiment. Consider, for instance, a capacitor-$\Phi$-memristor circuit subjected to a triangular voltage pulse~\cite{jkim19a}. Since the pulse amplitude can always be selected high enough to flip the magnetization while the voltage rises, and the voltage drop can always be selected slow enough to avoid the reverse flipping of magnetization, it is evident that the ``$\Phi$-memristor'' would fail the test since its final state would be different than the initial one, after the capacitor has fully discharged.  We emphasize that this conclusion can be reached based simply on the well-documented threshold property of magnetization reversal~\cite{Rudowicz03a}.

In conclusion, we have demonstrated that the ``$\Phi$-memristor'' Wang {\it et al.} claim to have ``discovered''~\cite{Wang19a} is, in fact, {\it not} a memristor. The fundamental error made by Wang {\it et al.} can be traced to their use of an {\it incorrect} model of magnetization dynamics. We have also highlighted several inconsistencies in their paper, including the inability to reproduce some of their curves, and their dubious design of a ``negative memristor'' emulator. Finally, the ``$\Phi$-memristor'' would fail the memristor test we have proposed, as a simple gedanken experiment shows. The current-carrying wire interacting with a magnetic core is nothing other than a memory inductor as we identified (together with Chua, co-author of Ref.~\cite{Wang19a}) in the past~\cite{diventra09a}.

\bibliography{litr}

\end{document}